\documentclass[aps,prl,superscriptaddress,amsmath,amssymb,amsfonts,twocolumn,showpacs]{revtex4-2}
\usepackage[english]{babel}

\usepackage[letterpaper,top=2cm,bottom=2cm,left=2.5cm,right=2.5cm,marginparwidth=1.75cm]{geometry}

\usepackage{amsmath,amsfonts,amssymb}
\usepackage{stmaryrd}
\usepackage{bm}
\usepackage{graphicx}
\usepackage[usenames,dvipsnames]{xcolor}
\usepackage[colorlinks=true, allcolors=blue]{hyperref}
\usepackage{ulem}

\usepackage{enumitem}
\setenumerate[1]{label=\arabic*)}

\newcommand{\orcid}[1]{\href{https://orcid.org/#1} 
  {\includegraphics[width=10pt]{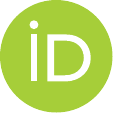}}}

 \newcommand{\para}[1]{ \noindent\textit{\textbf{#1---}}}

\newcommand{\e}{\eta}

\newcommand{\sech}{{\rm sech}}

\newcommand{\atanh}{{\rm atanh}}

\newcommand{\dd}{{\rm d}}

\begin{document}

\title{Exact Charge, Current, and Velocity Fields of Interacting Korteweg–de Vries Solitons}

\author{Thibault Bonnemain \orcid{0000-0003-0969-2413}} 
\affiliation{Laboratoire de Physique Théorique et Modélisation, CNRS UMR 8089,
CY Cergy Paris Université, 95302 Cergy-Pontoise Cedex, France}

\author{Fran\c{c}ois Copie \orcid{0000-0002-8617-2326}}
\affiliation{Univ. Lille, CNRS, UMR 8523, PhLAM – Physique des Lasers, Atomes et Molécules, F-59000 Lille, France}

\author{Pierre Suret \orcid{0000-0003-3527-4434}}
\affiliation{Univ. Lille, CNRS, UMR 8523, PhLAM – Physique des Lasers, Atomes et Molécules, F-59000 Lille, France}

\author{Stéphane Randoux \orcid{0000-0001-9309-6539}}
\affiliation{Univ. Lille, CNRS, UMR 8523, PhLAM – Physique des Lasers, Atomes et Molécules, F-59000 Lille, France}

\date{\small\today}

\begin{abstract}
Solitons in integrable systems exhibit a dual wave–particle character, yet their identification as individual objects becomes ambiguous during interactions, where they deform and delocalize. We develop a microscopic, field-based description that resolves this issue by introducing exact space–time fields of charge, current, and velocity derived from the inverse scattering transform (IST) of the Korteweg–de Vries (KdV) equation. This framework enables individual KdV solitons to be tracked throughout interactions, providing a quantitative description of their trajectories and deformation beyond the asymptotic regime. We show that the dynamics of $N$ interacting solitons can be formulated in terms of $N$ independent continuity equations, in which interactions are encoded in initial correlations that are subsequently propagated. From this microscopic viewpoint, effective velocities and hydrodynamic behavior emerge upon coarse graining, recovering kinetic theory of soliton gases and Generalized Hydrodynamics as scaling limits. Our results establish a direct connection between the wave-based IST formalism and particle-like emergent descriptions, offering a unified framework for soliton dynamics across scales.
\end{abstract}

\maketitle

Self-localized nonlinear wave packets known as solitons, sustained by an exact compensation between dispersion and nonlinearity, emerge as universal excitations across a wide range of physical platforms, from optics and quantum gases to fluids, metamaterials, and biological systems \cite{Rabec:25,Baker:25,Veenstra:25,Dieli:26,Lannig:20,Novkoski:22,Copie:23,Heimburg:05}. Solitons exhibit a dual nature in nonlinear physics, behaving both as particle-like entities and as coherent nonlinear waves. On the one hand, they display remarkable robustness, asymptotically recovering their shape, amplitude, and velocity after interactions, as first revealed in numerical simulations of the Korteweg–de Vries (KdV) equation by Zabusky and Kruskal \cite{Zabusky:65}. On the other hand, their wave nature is captured by the inverse scattering transform (IST), introduced by Gardner \textit{et al.} \cite{Gardner:67}, which provides an exact solution method for a broad class of integrable nonlinear dispersive equations \cite{Zakharov:72,Novikov_book,yang2010nonlinear,Ablowitz:73}.

A hallmark of soliton interactions is their elastic character: collisions preserve all intrinsic parameters and, asymptotically, result only in finite position shifts \cite{Novkoski:22,Copie:23}. Within the IST framework, solitons correspond to nonlinear normal modes associated with discrete eigenvalues that remain invariant in time \cite{Newell_book_solitons,Miura:76,Drazin_book}. However, during interactions, solitons deform and delocalize, obscuring their position, extent, and individual contribution to the wavefield \cite{fuchssteiner1987solitons,yoneyama1984interacting,nguyen2004soliton,benes2006decompositions,aggarwal2025asymptotic,aggarwal2025effective}. This raises a fundamental question: how can one define and track solitons as physical entities beyond the asymptotic (well-separated) regime? 

This issue becomes particularly acute in many-body settings, where solitons form coherent structures such as dispersive shock waves or statistical ensembles known as soliton gases (SGs) \cite{Maiden:18,EL:16,zakharov1971kinetic,GEl:05,Redor:19,Suret:20,Fache:24}. Effective descriptions based on kinetic theory \cite{zakharov1971kinetic,el2003thermodynamic,el2021soliton,GEl:20} and, more recently, Generalized Hydrodynamics (GHD) \cite{bertini2016transport,castro2016emergent,doyon2020lecture,spohn2024hydrodynamic,bonnemain2022generalized, koch2022generalized,Suret:24} successfully capture large-scale dynamics by treating solitons as interacting quasiparticles \cite{schemmer2019generalised,charnay2026experimental,Fache:25,kranzl2023observation}, thereby providing a framework to address fundamental questions related to, e.g., modulational instability or wave-mean flow interaction \cite{gelash2019bound,congy2026exactly}. Yet, these approaches rely on coarse-grained descriptions in which the underlying wave structure is implicit, leaving unclear what microscopic information is lost and how it affects dynamics at intermediate scales.

In this Letter, we develop a microscopic, field-based description of interacting solitons directly rooted in the IST spectral framework. Building on the association of the local phase-space density of KdV solitons with the squared eigenfunctions of the associated spectral problem \cite{Fache:25}, we construct exact space–time fields for soliton charge, current, and velocity. This representation provides a physically transparent picture in which individual solitons remain well defined even during strong interactions, beyond the standard isolated sech-profile description. Within this framework, effective velocities and local contributions to conserved quantities are naturally defined without invoking a collision-rate ansatz, and kinetic and hydrodynamic theories emerge as scaling limits. Our results establish an explicit bridge between the wave-based IST formalism and particle-based emergent descriptions, providing a unified view of soliton dynamics across scales.

\begin{figure*}[!]
  \includegraphics[width=1\textwidth]{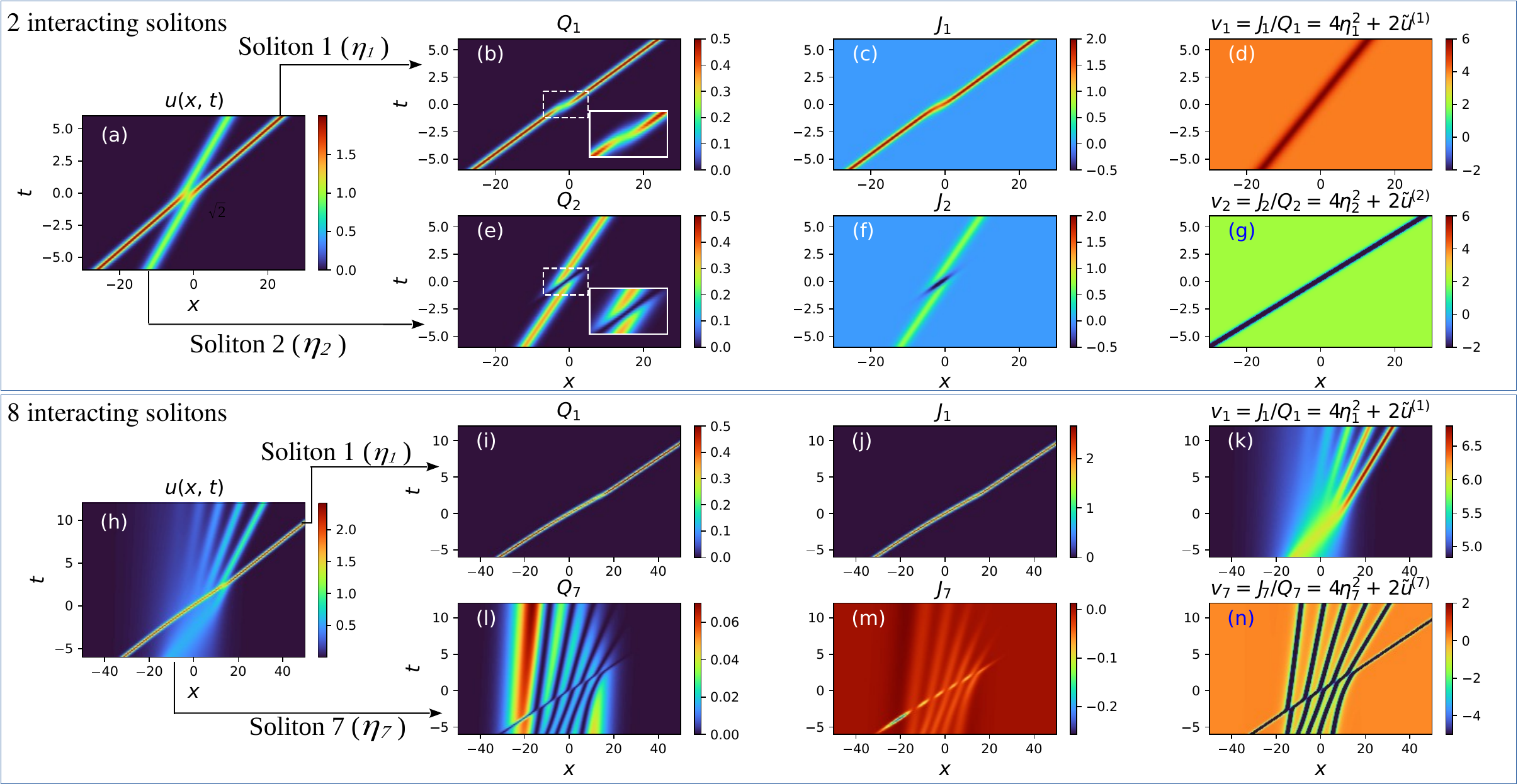}
  \vspace{-15pt}
  \caption{Numerical simulation showing the interaction between 2 (upper panel) and 8 (lower panel) KdV solitons. (a) The field $u(x,t)$ is computed from Eq. (\ref{eq:TraceFL}) while the eigenfunctions $\psi_{m}(x,t)$ are determined by solving the Marchenko system (\ref{eq:MarchSysMat}) using arbitrary numerical precision.  (b), (e), (i), (l)  Charge densities $Q_{m}(x,t)=\psi_{m}^2(x,t)$ associated with solitons with eigenvalues $\eta_{m}$. (c), (f), (j), (m)  Associated currents $J_{m}(x,t)$ computed using Eq. (\ref{eq:OnsagerJQ}). (d), (g), (k), (n) Associated velocity fields $v_{m}(x,t)$ computed using the definition $v_m=J_m/Q_m $ or equivalently the Darboux-deleted relation given by Eq. (\ref{eq:VelFieldDarDel}). In the upper panel, $\eta_1=1$, $\eta_2=1/\sqrt{2}$. In the lower panel, $\eta_m =[1.1, 0.7, 0.6, 0.5, 0.4, 0.3, 0.2, 0.1]$. }\label{fig:SolInt}
\end{figure*}

\para{IST formalism and basic ideas}
Because of its universal physical relevance, the KdV equation
\begin{equation}\label{eq:kdv}
u_t + 3(u^2)_x + u_{xxx} = 0 \; ,
\end{equation}
is seen as a paradigmatic example of nonlinear, dispersive, integrable partial differential equation featuring solitons \cite{ablowitz1981solitons,novikov1984theory,calogero2012integrability,remoissenet2013waves}. The KdV initial value problem is commonly addressed via the IST, by solving the auxiliary \textit{linear} problem \begin{equation}\label{Lpb} \psi_{xx} + (u+\lambda)\psi = 0 \; ,
\quad  \psi_t + \psi_{xxx} - 3(u+\lambda) \psi_x = 0 \; .
\end{equation}
If the KdV wavefield $u$ is purely solitonic, it can be reconstructed in terms of $N$ eigenfunctions $\psi_m$ and the associated invariant eigenvalues $\lambda_m = -\eta_m^2$ \cite{gardner1974korteweg}
\begin{equation}\label{eq:TraceFL}
    u(x,t) = 4\sum_{m=1}^N  \eta_m \psi_m^2(x,t) \; .
\end{equation}
All the eigenfunctions are real with normalisation $\int_{\mathbb R} \dd x\, \psi_m^2(x,t) = 1$, and are related to one another through the \textit{Marchenko system} of linear equations
    \begin{equation}\label{eq:MarchSysMat}
\mathcal A\,\bm{\psi}(x,t) = \bm{a}(x,t) \; .
\end{equation}
    Here $\bm{\psi}(x,t) := \big(\psi_1(x,t),\dots,\psi_N(x,t)\big)^{T}$ and $\bm{a} = \left(c_1(t)\exp(-\eta_1 x),  \ \cdots, \ c_n(t) \exp(-\eta_Nx)\right)^T$, while 
\begin{equation}
\mathcal A_{mn}(x,t)
:= \delta_{mn} + \frac{c_m(t)c_n(t)}{\e_m+\e_n}\,e^{-(\e_m+\e_n)x}\; ,
\end{equation}
where we introduced the \textit{norming constants} $c_m(t) = c_m(0)\exp\left(4\eta_m^3 t\right)$, which encode the initial condition.

A direct consequence of the decomposition \eqref{eq:TraceFL} is the identification of solitons with the squared eigenfunctions $Q_m := \psi_m^2$, which encode both their position and extension as they deform through interactions. This identification naturally reframes solitons as densities of (conserved) charges in motion. In this sense, the $Q_m$ provide a microscopic realization of the quasiparticle densities underlying GHD \cite{doyon2020lecture}, where the distribution of eigenvalues plays the role of a momentum distribution. This perspective is also consistent with recent rigorous approaches to the GHD of the Toda lattice \cite{aggarwal2025asymptotic,aggarwal2025effective}.

\para{Dynamics of the squared eigenfunctions}
 To study the dynamics of interacting solitons, we study that of the squared eigenfunctions. Multiplying the first equation of \eqref{Lpb} by $6\psi_x$, the second by $2\psi$, adding the results and using the identity $(\psi^2)_{xxx} = 3(\psi_x)^2_x+2\psi\psi_{xxx}$, we have \cite{gardner1974korteweg}
\begin{equation}\label{eq:ContQ}
    \partial_t Q_m + 6u\partial_x Q_m + \partial_{xxx} Q_m =0 \; ,
\end{equation}
sometimes called ``interacting KdV equations'' \cite{yoneyama1984interacting}. 

We propose a new interpretation of these equations and show they can be put in more convenient forms which do not involve the wavefield $u$. Because of their normalisation, the $Q_m$'s are, by definition, conserved densities, so that there exists $N$ associated currents, $J_m$'s, defined by
\begin{equation}\label{eq:ContEqEigen}
    \partial_t Q_m + \partial_x J_m = 0 \; ,
\end{equation}
with $J_m = \int_{-\infty}^x \dd x \left[6u\partial_x Q_m + \partial_{xxx}Q_m\right]$. To explicitly compute these currents, recall the auxiliary problem \eqref{Lpb} which implies $6u\partial_x Q_m = - 6\partial_x\left[(\partial_x\psi_m)^2 + \lambda Q\right]$,
hence
\begin{equation}\label{eq:OnsagerJQ}
    J_m = -6\left(\partial_x\sqrt{Q_m}\right)^2 + 6 \eta_m^2Q_m + \partial_{xx} Q_m \; ,
\end{equation}
meaning we may define a velocity field $v_m$ that only depends on the associated $Q_m$
\begin{equation}\label{eq:VelField}
    v_m := \frac{J_m}{Q_m} = -\frac{3}{2}\left(\partial_x\log Q_m\right)^2 + \frac{\partial_{xx}Q_m}{Q_m} + 6\eta_m^2 \; .
\end{equation}
This explicit relation between $v_m$ and $Q_m$ is a central result of this work and can be viewed as a ``{\it microscopic equation of state}". It shows that solitons effectively decouple: the evolution of each $Q_m$ is independent of other solitons, interactions being solely encoded in the initial conditions through the Marchenko system \eqref{eq:MarchSysMat}; initial correlations are merely propagated. This picture is consistent with recent developments in GHD \cite{hubner2025diffusive,hubner2026diffusive} and provides a new perspective on diffusion from convection \cite{medenjak2020diffusion} in integrable PDEs.

Importantly, Eq.~\eqref{eq:ContEqEigen} provides a simple way to follow and visualize the dynamics of a given soliton (at which speeds it moves \textit{and} how it deforms) through a physically enlightening interpretation in terms of charge and current densities. This is illustrated in Fig.~\ref{fig:SolInt}, which shows the charge, current, and velocity fields associated with the interaction of two (upper panel) and eight (lower panel) KdV solitons. For two interacting solitons, the charge fields $Q_m$ remain localized as $\sech^2$ pulses [Figs.~\ref{fig:SolInt}(b), \ref{fig:SolInt}(e)], although the slower soliton briefly loses weight near the collision point. The current fields $J_m$ follow similar space–time dynamics [Figs.~\ref{fig:SolInt}(c), \ref{fig:SolInt}(f)]. The velocity field $v_m=J_m/Q_m$ of a given soliton [Figs.~\ref{fig:SolInt}(d), \ref{fig:SolInt}(g)] appears as a uniform background of amplitude $4\eta_m^2$ (the free velocity of an isolated KdV soliton with eigenvalue $\eta_m$), locally perturbed by the presence of the other soliton. Exact analytical expressions for the fields in binary collisions can be obtained within the framework developed below and are detailed in End Matter.

Fig.~\ref{fig:SolInt}(h) shows a tracer soliton with $\eta_1=1.1$ interacting with a dense ensemble of seven solitons. Its charge and current fields remain sharply localized throughout the evolution [Figs.~\ref{fig:SolInt}(i),~\ref{fig:SolInt}(j)], allowing it to be unambiguously tracked, whereas a strongly interacting soliton with $\eta_7=0.2$ is initially markedly delocalized [Fig.~\ref{fig:SolInt}(l)] before gradually relocalizing at long times under the KdV evolution. The weaker localization of smaller-amplitude solitons during interactions follows naturally from the auxiliary problem \eqref{Lpb} (see Sturm’s oscillation theorem \cite{simon2005sturm} and Ref.~\cite{johnson1982rotation}). As before, the velocity fields consist of a uniform background $4\eta_m^2$, locally perturbed by interactions with other solitons [Figs.~\ref{fig:SolInt}(k),~\ref{fig:SolInt}(n)].

\para{Alternative form of the velocity field} Although Eqs.~\eqref{eq:ContEqEigen}-\eqref{eq:VelField} are physically interesting and numerically practical, they are not the best suited to extract analytical results. We now derive an alternative expression of the velocity field \eqref{eq:VelField} which may be more convenient to deal with. Via the auxiliary problem \eqref{Lpb}, one may write
\begin{equation}\label{eq:Velfield2}
\begin{aligned}
    v_m =8\e_m^2 -2 u
- 4\left(\partial_x\log\psi_m\right)^2 \; .
\end{aligned}
\end{equation}
We now introduce the  \textit{Darboux-deleted potential} \cite{sukumar2004supersymmetric,gu2004darboux}
\begin{equation}\label{eq:DarDel}
    \tilde u^{(m)} = u + 2\partial_{xx}\log\psi_m \; ,
\end{equation}
corresponding to the purely solitonic solution $u$ in which the $m-$th soliton has been removed. To be more precise, consider the determinant representation of the wavefield $u(x,t) = 2\partial_{xx}\log \left[ \det \mathcal A(x,t)\right]$  \cite{kay1956reflectionless,hirota2004direct}, which connects directly to the Marchenko system \eqref{eq:MarchSysMat}. Any eigenfunction can be written as a ratio of determinants via Cramer's rule, $\psi_m(x,t)={\det \mathcal A^{(m)}(x,t)}/{\det \mathcal A(x,t)}$, in which the matrix $\mathcal A^{(m)}$ is obtained from $\mathcal A$ by replacing its $m$-th
column by the vector $\bm{a}$, so that
\begin{equation}\label{eq:DarDet}
    \tilde u^{(m)} = 2\partial_{xx} \log \left[\det\mathcal A^{(m)}(x,t)\right] \; .
\end{equation}
Combining Eqs.~\eqref{eq:Velfield2} and \eqref{eq:DarDel} eventually yields
\begin{equation}\label{eq:VelFieldDarDel}
\begin{aligned}
    v_m(x,t) &= 4\e_m^2 + 2\,\tilde u^{(m)}(x,t) \, ,
\end{aligned}
\end{equation}
the velocity field $v_m$ of a soliton is given by its free velocity $4\e_m^2$, locally renormalized by the field $\tilde u^{(m)}$ generated by all other solitons composing $u$, and explicitly constructed through Eq.~\eqref{eq:DarDet}. Note that  $\tilde u^{(m)} \neq 4\sum_{j\neq m}
\eta_j \psi_j^2$ because of the nonlinear interactions between solitons: removing an eigenfunction from system \eqref{eq:MarchSysMat} clearly affects all the others.

Eq.~\eqref{eq:VelFieldDarDel} allows to reformulate the continuity equation \eqref{eq:ContEqEigen} as a \textit{linear} PDE solvable via the method of characteristics \cite{courant2024methods}. Here, characteristic curves are defined by 
\begin{equation}\label{eq:Charac}
    \dot  x_m(t,\bar x_m) \equiv \frac{\dd x_m(t,\bar x_m)}{\dd t} = v_m(x_m(t,\bar x_m),t) \; ,
\end{equation}
they correspond to trajectories of point particles starting at position $\bar x_m$ moving within the velocity field $v_m(x,t)$. Along those characteristics, the squared eigenfunctions evolve according to 
\begin{equation}\label{eq:CharacQ}
    \dot Q_m(x_m(t, \bar x_m),t) = -Q_m(x,t) \partial_x v_m(x,t)|_{x=x_m(t,\bar x_m)} \; ,
\end{equation}
with initial condition $Q_m(x,t=0) = \bar Q_m(x)$. A detailed analytical treatment of binary KdV soliton collisions, based on Eqs.~\eqref{eq:Charac}–\eqref{eq:CharacQ}, is provided in End Matter and accurately reproduces the trajectory in Fig.~\ref{fig:SolInt}(b), naturally recovering the soliton position shift. Extensions to rarefied SGs and to a particular type of SG known as \textit{genus 1 condensate} are also discussed.

\para{Kinetic equation and soliton gas theory} The charge, current, and velocity fields introduced above provide new insight into the emergent large-scale dynamics of integrable many-body systems. SG theory and GHD are effective field theories, describing dynamics at scales $X=\epsilon x$ and $T=\epsilon t$ much larger than those associated with individual solitons $(\epsilon \ll 1)$. They deal with the \textit{density of states} (DOS) $\rho$, defined such that $\rho(\eta;X,T)\dd X \dd\eta$ gives the number of solitons in the phase-space region $[X;X+\dd X]\times[\eta;\eta+\dd\eta]$ at time $T$. Physical observables are evaluated through the DOS \cite{el2021soliton,doyon2020lecture,bonnemain2022generalized}.

Within SG kinetic theory and GHD, the large-scale evolution of $\rho$ is governed by
\begin{equation}\label{eq:GHDeq}
\partial_T \rho (\eta;X,T) + \partial_X\left[v^{{\rm eff}}(\eta;X,T)\rho(\eta;X,T)\right] = 0 , 
\end{equation}
where the effective velocity satisfies
\begin{equation}\label{eq:Veff}
\begin{aligned}
&v^{{\rm eff}}(\eta) = 4\eta^2 + \int_\Gamma \dd\mu\, \phi(\eta,\mu) \rho(\mu)[v^{{\rm eff}}(\eta)-v^{{\rm eff}}(\mu)]  ,
\end{aligned}
\end{equation}
with $\phi(\eta,\mu)=1/\eta \log ( |(\eta + \mu) / (\eta - \mu)| )$. Eq.~\eqref{eq:Veff} admits an interpretation in terms of a collision-rate ansatz \cite{el2021soliton}: on large scales, solitons behave as point particles undergoing instantaneous shifts $\phi(\eta,\mu)$ upon collisions, leading to a renormalized velocity. By contrast, the velocity field defined by Eq.~\eqref{eq:VelField} retains information about the spatial structure and deformation of solitons during interactions, and therefore cannot be directly identified with the effective velocity of SG theory. The latter instead emerges from a reduced description in which such microscopic details are averaged out.

We now introduce a microscopic definition of the DOS and of the effective velocity. Local averages of the field are given by \cite{GEl:21,congy2023dispersive,congy2024statistics}
\begin{equation}\label{eq:EnsembleAv}
\frac{1}{2l_\epsilon}\int_{\frac{X}{\epsilon} - l_\epsilon}^{\frac{X}{\epsilon} + l_\epsilon} \dd x\, u(x,T\epsilon^{-1}) = 4\int_\Gamma \dd\eta\, \eta\rho\left(\eta;X,T\right) \; ,
\end{equation}
where $\epsilon \ll 1$, $\lim_{\epsilon \to 0} l_\epsilon \to \infty$, and $\lim_{\epsilon \to 0} \epsilon l_\epsilon \to 0$. Using decomposition \eqref{eq:TraceFL}, as in ref.~\cite{Fache:25}, we interpret $\rho$ as interpolating between local averages of squared eigenfunctions,
\begin{equation}\label{eq:DOS}
\rho(\eta_m;X,T) \equiv \frac{1}{2l_\epsilon}\int_{\frac{X}{\epsilon} - l_\epsilon}^{\frac{X}{\epsilon} + l_\epsilon} \dd x\, Q_m(x,T\epsilon^{-1})  .
\end{equation}
This motivates the \textit{empirical} (in the sense of mathematical statistics \cite{hult2016large,withers2010distribution}), \textit{local} definition of the DOS
\begin{equation}\label{eq:EmpDOS}
\tilde \rho(\eta;x,t) = \sum_{m=1}^N \delta(\eta-\eta_m)Q_m(x,t)  .
\end{equation}
For random soliton ensembles, $\tilde \rho$ should self-average to Eq.~\eqref{eq:DOS} when sufficiently many solitons are contained in an interval of size $2l_\epsilon$ \cite{Fache:25}. A detailed analysis of convergence in $\epsilon$ and $l_\epsilon$, as well as the equivalence between spatial, temporal, and ensemble averages, is left for future work. 

Introducing the empirical spectral flux density $\tilde f(\eta;x,t)=\sum_{m=1}^N \delta(\eta-\eta_m), v_m(x,t) Q_m(x,t)$, we derive a continuity equation for $\tilde \rho$ and perform a mesoscopic spatiotemporal average, yielding
\begin{equation}\label{eq:MicroGHDeq_interm}
\begin{aligned}
&\frac{1}{4 l_\epsilon \tau_\epsilon}\int_{\frac{X}{\epsilon}-l_\epsilon}^{\frac{X}{\epsilon}+l_\epsilon} \dd x\int_{\frac{T}{\epsilon}-\tau_\epsilon}^{\frac{T}{\epsilon}+\tau_\epsilon} \dd t \left[ \partial_t \tilde\rho(\eta;x,t) + \partial_x \tilde f(\eta;x,t)\right]  \\
&= 0,
\end{aligned}
\end{equation}
where $\tau_\epsilon$ is the temporal analogue of $l_\epsilon$. Eq.~\eqref{eq:MicroGHDeq_interm} simplifies to
\begin{equation}\label{eq:MicroGHDeq}
\begin{aligned}
\partial_T \rho(\eta;X,T) + \partial_X\left( \frac{1}{2\tau_\epsilon}\int_{\frac{T}{\epsilon}-\tau_\epsilon}^{\frac{T}{\epsilon}+\tau_\epsilon} \dd t\tilde f(\eta;x,t)\right) = 0  .
\end{aligned}
\end{equation}
As such, the effective velocity~\eqref{eq:Veff} arises from a reduction of degrees of freedom, with the second term in \eqref{eq:GHDeq} approximating that in \eqref{eq:MicroGHDeq} at large scales.

\begin{figure}[h]
\includegraphics[width=0.5\textwidth]{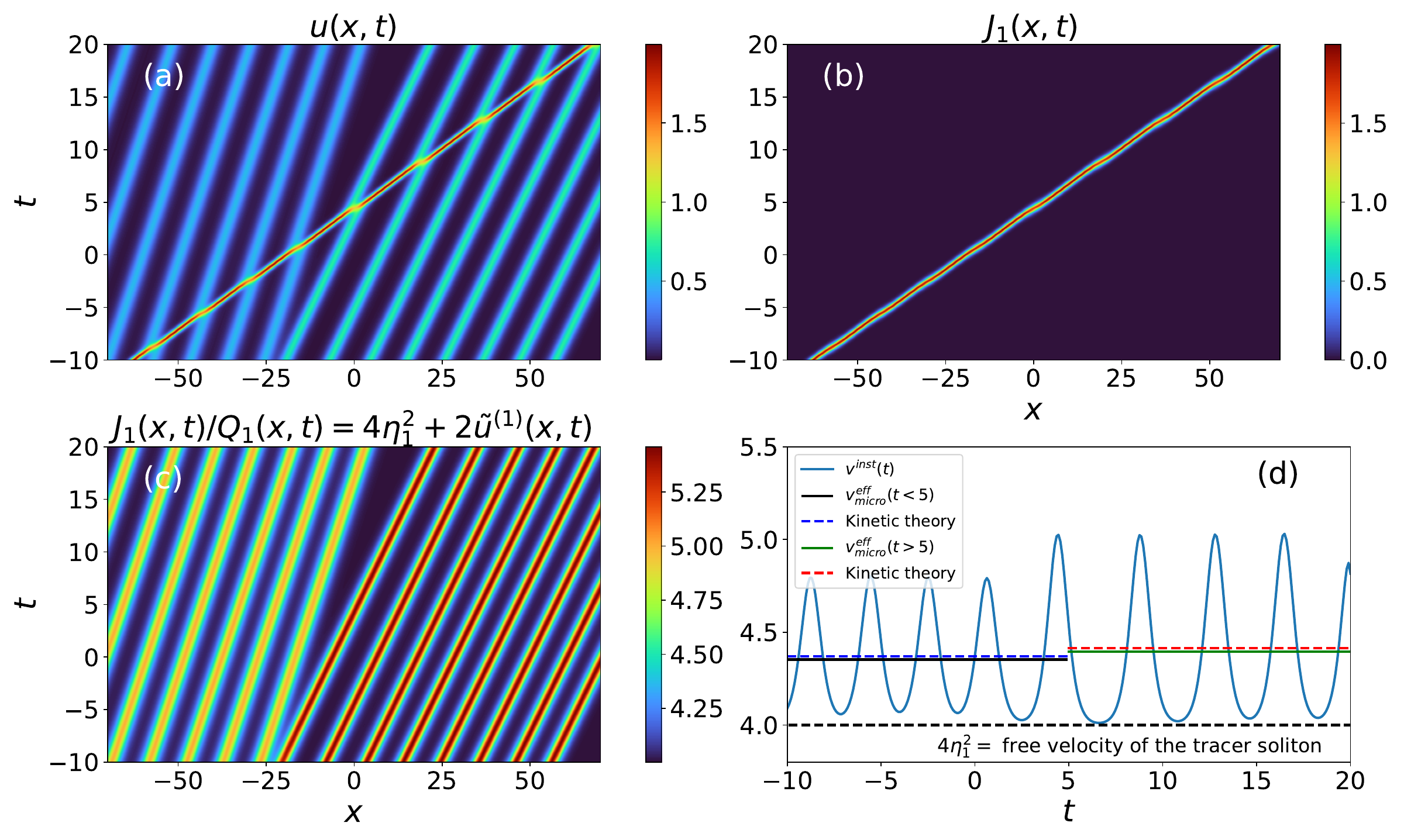}
\vspace{-15pt}
\caption{Numerical simulation showing the interaction of a tracer soliton with a bichromatic dilute SG composed of two noninteracting species. (a) KdV field $u(x,t)$. (b) Current field $J_1(x,t)$ of the tracer soliton with eigenvalue $\eta_1=1$. (c) Velocity field $v_1(x,t)=J_1/Q_1$ of the tracer soliton. (d) Time evolution of the instantaneous velocity $v^{\rm inst}_{1}$ (Eq. \eqref{eq:VeffMicro}) of the tracer soliton. The mean value of the fluctuations, $v^{\rm eff}_{{\rm micro}}$, coincide with prediction of the kinetic theory (given by Eq. (10) of Ref. \cite{Carbone:16}), as indicated by horizontal full and dashed lines.
}\label{fig:velocities}
\end{figure}

While a proper derivation of KdV soliton hydrodynamics is left for future work, we focus here on a practical microscopic definition of the effective velocity. We define it as the averaged velocity of the soliton center of mass $\langle x\rangle_j(t) := \int_{\mathbb R} x Q_j(x,t)\dd x$, whose trajectory follows a characteristic \eqref{eq:Charac}. We thus introduce the instantaneous velocity as
\begin{equation}\label{eq:VInst}
\begin{aligned}
& v^{\rm inst}_{j}(t) = \partial_t \langle x\rangle_j(t) = \int_{\mathbb R} \dd x \ J_j(x,t) ,
\end{aligned}
\end{equation}
where we used that $xQ_j \to 0$ as $|x|\to\infty$. This definition follows from the first Ehrenfest theorem applied to the auxiliary Schrödinger problem \eqref{Lpb} \cite{hall2013quantum}, and is consistent with recent microscopic approaches to the Toda lattice \cite{aggarwal2025asymptotic}. From this, we define the ``microscopic" effective velocity as
\begin{equation}\label{eq:VeffMicro}
\begin{aligned}
& v^{\rm eff}_{{\rm micro}}(\e_j;\langle x\rangle_j(T),T) = \frac{1}{2\tau_\epsilon}\int_{\frac{T}{\epsilon}-\tau_\epsilon}^{\frac{T}{\epsilon}+\tau_\epsilon} \dd t\, v^{\rm inst}_{j}(t)  .
\end{aligned}
\end{equation}
The physical meaning of the microscopic effective velocity is illustrated in Fig.~\ref{fig:velocities}, which shows the interaction of a tracer soliton with a bichromatic dilute SG. The tracer has eigenvalue $\eta_1=1$, while the DOS of the SG is $\rho(\eta)=\rho_2 \delta(\eta-\eta_2)+\rho_3 \delta(\eta-\eta_3)$ with $\eta_2=0.5$ and $\eta_3=0.6$. Figs.~\ref{fig:velocities}(a)–(c) display the KdV field $u$, the current $J_1$ and velocity $v_1=J_1/Q_1$ fields of the tracer, while Fig.~\ref{fig:velocities}(d) shows the time evolution of $v^{\rm inst}_{1}$ computed from Eq.~\eqref{eq:VInst}. The instantaneous velocity of the tracer soliton exhibits fluctuations due to collision with solitons in the SG, with amplitudes set by the spectral mismatch $|\eta_1-\eta_{2,3}|$. Its mean value $v^{\rm eff}_{{\rm micro}}$ (full horizontal lines) closely matches the kinetic-theory prediction (dashed horizontal lines) obtained from Ref.~\cite{Carbone:16}. Beyond reproducing the average behavior given by the kinetic theory, this demonstrates that our approach captures the velocity fluctuations. However, our approach is not limited to such simple diluted SGs, e.g it provides a more general, direct and transparent route to results analogous to those of \cite{girotti2023soliton} (see End Matter), where the velocity of a tracer soliton propagating through a deterministic soliton gas was shown to agree with Eq.~\eqref{eq:Veff}.

\para{Discussion} We discussed a novel and physically enlightening interpretation of the dynamics of $N$ interacting solitons in terms of $N$ independent continuity equations \eqref{eq:ContEqEigen}. The definition \eqref{eq:VelField} of a velocity \textit{field} allows to clearly follow the trajectory of a soliton and the way it deforms along it. The independent equations \eqref{eq:ContEqEigen} can further be linearized, and then solved via the method of characteristics, by introducing the Darboux-deleted potential which gives rise to a physically meaningful interpretation of the velocity field. Our interpretation of the microscopic dynamics of interacting solitons paves the way for the derivation of soliton hydrodynamics from first principles, suggesting a natural form for the empirical density \eqref{eq:EmpDOS} and for the effective velocity \eqref{eq:VeffMicro}. Moreover, knowledge of the exact microscopic charge, current and velocity fields, will provide a way to not only extract expected values of observables, but fluctuations as well, going beyond SG theory.

A complete derivation of the kinetic equation \eqref{eq:GHDeq} nevertheless requires a better understanding of soliton localization in the thermodynamic limit. While eigenfunctions remain localized for finite $N$, localization becomes nontrivial in SGs as $N\to\infty$ and the support of $u$ diverges. In particular, the genus-0 soliton condensate corresponds almost surely to a constant field satisfying $\forall j,\ \e_j^2-u(x,t)\leq0$ \cite{congy2023dispersive}, implying complete delocalization of all solitons. This suggests that localization properties depend strongly on the underlying state. Determining which solitons remain dynamically relevant in dense random gases is therefore an important open problem, partially addressed in \cite{doyon2026where}. 

Lastly, we must stress that this work finds its impetus in ref.~\cite{Fache:25} where SG theory and decomposition \eqref{eq:TraceFL} were used to analyze the propagation of a complex signal through a nonlinear electrical line. This was despite the fact that such propagation was described by the nonintegrable KdV-Burgers equation. Our approach can naturally be extended to the context of \cite{Fache:25}, highlighting the fact it is not restricted to the KdV (nor to an integrable) equation.

\para{Data availability} Data will be provided upon reasonable requests.

\begin{acknowledgments} 
\para{Acknowledgments}  This work was partially supported by the Agence Nationale de la Recherche through the SOGOOD (Grant No. ANR-21-CE30-0061) project. We acknowledge the support of the CDP C2EMPI, as well as the French State under the France-2030 programme, the University of Lille, the Initiative of Excellence of the University of Lille, the European Metropolis of Lille for their funding and support of the R-CDP-24-004-C2EMPI project.
  The authors thank Benjamin Doyon, Gennady El and Tamara Grava for fruitful discussions. 
\end{acknowledgments}

\bibliographystyle{ieeetr}
\bibliography{Biblio,Biblio-2019-01-03,Bibliography,SG_sr}

@book{novikov1984theory,
  title={Theory of solitons: the inverse scattering method},
  author={Novikov, S and Manakov, SV and Pitaevskii, LP and Zakharov, Vladimir E},
  year={1984},
  publisher={Springer Science \& Business Media}
}

@book{ablowitz1981solitons,
  title={Solitons and the inverse scattering transform},
  author={Ablowitz, Mark J and Segur, Harvey},
  year={1981},
  publisher={SIAM}
}

@article{bertini2016transport,
  title={Transport in out-of-equilibrium x x z chains: Exact profiles of charges and currents},
  author={Bertini, Bruno and Collura, Mario and De Nardis, Jacopo and Fagotti, Maurizio},
  journal={Physical review letters},
  volume={117},
  number={20},
  pages={207201},
  year={2016},
  publisher={APS}
}

@article{benes2006decompositions,
  title={On decompositions of the KdV 2-Soliton},
  author={Benes, N and Kasman, A and Young, K},
  journal={Journal of nonlinear science},
  volume={16},
  number={2},
  pages={179--200},
  year={2006},
  publisher={Springer Alemania}
}

@book{hall2013quantum,
  title={Quantum theory for mathematicians},
  author={Hall, Brian C},
  year={2013},
  publisher={Springer}
}

@book{remoissenet2013waves,
  title={Waves called solitons: concepts and experiments},
  author={Remoissenet, Michel},
  year={2013},
  publisher={Springer Science \& Business Media}
}

@article{fuchssteiner1987solitons,
  title={Solitons in interaction},
  author={Fuchssteiner, Benno},
  journal={Progress of Theoretical Physics},
  volume={78},
  number={5},
  pages={1022--1050},
  year={1987},
  publisher={Oxford University Press}
}

@book{spohn2024hydrodynamic,
  title={Hydrodynamic scales of integrable many-body systems},
  author={Spohn, Herbert},
  year={2024},
  publisher={World Scientific}
}

@article{congy2026exactly,
  title={Exactly Solvable Model of Wave-Mean Field Interaction in Integrable Turbulence},
  author={Congy, T and El, GA and Hoefer, MA},
  journal={Physical Review Letters},
  volume={136},
  number={14},
  pages={147201},
  year={2026},
  publisher={APS}
}

@inproceedings{doyon2026generalised,
  title={Generalised TT{\={}}-Deformations of Classical Free Particles: B. Doyon et al.},
  author={Doyon, Benjamin and H{\"u}bner, Friedrich and Yoshimura, Takato},
  booktitle={Annales Henri Poincar{\'e}},
  pages={1--50},
  year={2026},
  organization={Springer}
}

@book{byrd2013handbook,
  title={Handbook of elliptic integrals for engineers and physicists},
  author={Byrd, Paul F and Friedman, Morris D},
  year={2013},
  publisher={Springer}
}

@article{aggarwal2025effective,
  title={Effective velocities in the Toda lattice},
  author={Aggarwal, Amol},
  journal={Communications on Pure and Applied Mathematics},
  pages={e70046},
  year={2025},
  publisher={Wiley Online Library}
}

@article{kranzl2023observation,
  title={Observation of magnon bound states in the long-range, anisotropic Heisenberg model},
  author={Kranzl, Florian and Birnkammer, Stefan and Joshi, Manoj K and Bastianello, Alvise and Blatt, Rainer and Knap, Michael and Roos, Christian F},
  journal={Physical Review X},
  volume={13},
  number={3},
  pages={031017},
  year={2023},
  publisher={APS}
}

@article{charnay2026experimental,
  title={Experimental observation of ballistic correlations in integrable turbulence},
  author={Charnay, Elias and Escoubet, Adrien and Copie, Francois and Randoux, Stephane and Bonnemain, Thibault and Bastianello, Alvise and Suret, Pierre},
  journal={arXiv preprint arXiv:2601.21085},
  year={2026}
}

@article{nguyen2004soliton,
  title={Soliton collisions and ghost particle radiation},
  author={Nguy{\^e}{\~n}, Hi{\^e}{\'u} D},
  journal={Journal of nonlinear mathematical physics},
  volume={11},
  number={2},
  pages={180--198},
  year={2004},
  publisher={Springer}
}

@incollection{simon2005sturm,
  title={Sturm oscillation and comparison theorems},
  author={Simon, Barry},
  booktitle={Sturm-Liouville Theory: Past and Present},
  pages={29--43},
  year={2005},
  publisher={Springer}
}

@article{kay1956reflectionless,
  title={Reflectionless transmission through dielectrics and scattering potentials},
  author={Kay, I and Moses, HE},
  journal={Journal of Applied Physics},
  volume={27},
  number={12},
  pages={1503--1508},
  year={1956},
  publisher={American Institute of Physics}
}

@article{johnson1982rotation,
  title={The rotation number for almost periodic potentials},
  author={Johnson, Russell and Moser, J{\"u}rgen},
  journal={Communications in Mathematical Physics},
  volume={84},
  number={3},
  pages={403--438},
  year={1982},
  publisher={Springer}
}

@article{yoneyama1984interacting,
  title={Interacting Korteweg-de Vries equations and attractive Soliton interaction},
  author={Yoneyama, Tohru},
  journal={Progress of theoretical physics},
  volume={72},
  number={6},
  pages={1081--1088},
  year={1984},
  publisher={Oxford University Press}
}

@article{aggarwal2025asymptotic,
  title={Asymptotic scattering relation for the Toda lattice},
  author={Aggarwal, Amol},
  journal={arXiv preprint arXiv:2503.08018},
  year={2025}
}

@article{hubner2025diffusive,
  title={Diffusive hydrodynamics from long-range correlations},
  author={H{\"u}bner, Friedrich and Biagetti, Leonardo and De Nardis, Jacopo and Doyon, Benjamin},
  journal={Physical Review Letters},
  volume={134},
  number={18},
  pages={187101},
  year={2025},
  publisher={APS}
}

@article{hubner2026diffusive,
  title={Diffusive hydrodynamics of hard rods from microscopics},
  author={H{\"u}bner, Friedrich and Biagetti, Leonardo and De Nardis, Jacopo and Doyon, Benjamin},
  journal={SciPost Physics Core},
  volume={9},
  number={1},
  pages={010},
  year={2026}
}

@article{medenjak2020diffusion,
  title={Diffusion from convection},
  author={Medenjak, Marko and De Nardis, Jacopo and Yoshimura, Takato},
  journal={SciPost Physics},
  volume={9},
  number={5},
  pages={075},
  year={2020}
}

@book{gu2004darboux,
  title={Darboux transformations in integrable systems: theory and their applications to geometry},
  author={Gu, Chaohao and Hu, Hesheng and Zhou, Zixiang},
  year={2004},
  publisher={Springer Science \& Business Media}
}

@inproceedings{sukumar2004supersymmetric,
  title={Supersymmetric quantum mechanics and its applications},
  author={Sukumar, CV},
  booktitle={AIP Conference Proceedings},
  volume={744},
  number={1},
  pages={166--235},
  year={2004},
  organization={American Institute of Physics}
}

@article{girotti2023soliton,
  title={Soliton versus the gas: Fredholm determinants, analysis, and the rapid oscillations behind the kinetic equation},
  author={Girotti, Manuela and Grava, Tamara and Jenkins, Robert and McLaughlin, Ken T-R and Minakov, Alexander},
  journal={Communications on Pure and Applied Mathematics},
  volume={76},
  number={11},
  pages={3233--3299},
  year={2023},
  publisher={Wiley Online Library}
}

@article{gelash2019bound,
  title={Bound state soliton gas dynamics underlying the spontaneous modulational instability},
  author={Gelash, Andrey and Agafontsev, Dmitry and Zakharov, Vladimir and El, Gennady and Randoux, St{\'e}phane and Suret, Pierre},
  journal={Physical review letters},
  volume={123},
  number={23},
  pages={234102},
  year={2019},
  publisher={APS}
}

@article{zakharov1971kinetic,
  title={Kinetic equation for solitons},
  author={Zakharov, VE},
  journal={Sov. Phys. JETP},
  volume={33},
  number={3},
  pages={538--540},
  year={1971}
}

@article{hult2016large,
  title={Large deviations for weighted empirical measures arising in importance sampling},
  author={Hult, Henrik and Nyquist, Pierre},
  journal={Stochastic Processes and their Applications},
  volume={126},
  number={1},
  pages={138--170},
  year={2016},
  publisher={Elsevier}
}

@article{koch2022generalized,
  title={Generalized hydrodynamics of the attractive non-linear Schr\"odinger equation},
  author={Koch, Rebekka and Caux, Jean-S{\'e}bastien and Bastianello, Alvise},
  journal={Journal of Physics A: Mathematical and Theoretical},
  volume={55},
  number={13},
  pages={134001},
  year={2022},
  publisher={IOP Publishing}
}

@article{withers2010distribution,
  title={The distribution and quantiles of functionals of weighted empirical distributions when observations have different distributions},
  author={Withers, Christopher S and Nadarajah, Saralees},
  journal={Statistics \& probability letters},
  volume={80},
  number={13-14},
  pages={1093--1102},
  year={2010},
  publisher={Elsevier}
}

@book{courant2024methods,
  title={Methods of mathematical physics, volume 2},
  author={Courant, Richard and Hilbert, David},
  year={2024},
  publisher={John Wiley \& Sons}
}

@article{congy2024statistics,
  title={Statistics of extreme events in integrable turbulence},
  author={Congy, T and El, GA and Roberti, G and Tovbis, A and Randoux, S and Suret, P},
  journal={Physical Review Letters},
  volume={132},
  number={20},
  pages={207201},
  year={2024},
  publisher={APS}
}

@book{calogero2012integrability,
  title={What is integrability?},
  author={Calogero, Francesco and Ercolani, N and Flaschka, H and Marchenko, VA and Mikhailov, AV and Newell, AC and Schulman, EI and Shabat, AB and Siggia, ED and Sokolov, VV and others},
  year={2012},
  publisher={Springer Science \& Business Media}
}

@article{bonnemain2022generalized,
  title={Generalized hydrodynamics of the KdV soliton gas},
  author={Bonnemain, Thibault and Doyon, Benjamin and El, Gennady},
  journal={Journal of Physics A: Mathematical and Theoretical},
  volume={55},
  number={37},
  pages={374004},
  year={2022},
  publisher={IOP Publishing}
}

@article{doyon2024new,
  title={New classical integrable systems from generalized TT-deformations},
  author={Doyon, Benjamin and H{\"u}bner, Friedrich and Yoshimura, Takato},
  journal={Physical Review Letters},
  volume={132},
  number={25},
  pages={251602},
  year={2024},
  publisher={APS}
}

@misc{doyon2026where,
      title={Where solitons are in a KdV soliton gas}, 
      author={Benjamin Doyon},
      year={2026},
      eprint={2605.18093},
      archivePrefix={arXiv},
      primaryClass={math-ph},
      url={https://arxiv.org/abs/2605.18093}, 
}

@book{hirota2004direct,
  title={The direct method in soliton theory},
  author={Hirota, Ryogo},
  number={155},
  year={2004},
  publisher={Cambridge University Press}
}

@article{gardner1974korteweg,
  title={Korteweg-devries equation and generalizations. VI. methods for exact solution},
  author={Gardner, Clifford S and Greene, John M and Kruskal, Martin D and Miura, Robert M},
  journal={Communications on pure and applied mathematics},
  volume={27},
  number={1},
  pages={97--133},
  year={1974},
  publisher={Wiley Online Library}
}

@article{congy2023dispersive,
  title={Dispersive hydrodynamics of soliton condensates for the Korteweg--de Vries equation},
  author={Congy, T and El, GA and Roberti, G and Tovbis, A},
  journal={Journal of Nonlinear Science},
  volume={33},
  number={6},
  pages={104},
  year={2023},
  publisher={Springer}
}

@article{el2003thermodynamic,
title = "The thermodynamic limit of the Whitham equations",
journal = "Physics Letters A",
volume = "311",
number = "4",
pages = "374 - 383",
year = "2003",
issn = "0375-9601",
doi = "https://doi.org/10.1016/S0375-9601(03)00515-2",
url = "http://www.sciencedirect.com/science/article/pii/S0375960103005152",
author = "G.A. El"
}

@article{castro2016emergent,
  title={Emergent hydrodynamics in integrable quantum systems out of equilibrium},
  author={Castro-Alvaredo, Olalla A and Doyon, Benjamin and Yoshimura, Takato},
  journal={Physical Review X},
  volume={6},
  number={4},
  pages={041065},
  year={2016},
  publisher={APS}
}

@article{doyon2020lecture,
  title={Lecture notes on generalised hydrodynamics},
  author={Doyon, Benjamin},
  journal={SciPost Physics Lecture Notes},
  pages={018},
  year={2020}
}

@article{schemmer2019generalised,
  title = {Generalized Hydrodynamics on an Atom Chip},
  author = {Schemmer, M. and Bouchoule, I. and Doyon, B. and Dubail, J.},
  journal = {Phys. Rev. Lett.},
  volume = {122},
  issue = {9},
  pages = {090601},
  numpages = {7},
  year = {2019},
  month = {Mar},
  publisher = {American Physical Society},
  doi = {10.1103/PhysRevLett.122.090601},
  url = {https://link.aps.org/doi/10.1103/PhysRevLett.122.090601}
}

@article{el2021soliton,
  title={Soliton gas in integrable dispersive hydrodynamics},
  author={El,  Gennady A},
  journal={Journal of Statistical Mechanics: Theory and Experiment},
  volume={2021},
  number={11},
  pages={114001},
  year={2021},
  publisher={IOP Publishing}
}

@article{Lannig:20,
  title = {Collisions of Three-Component Vector Solitons in Bose-Einstein Condensates},
  author = {Lannig, Stefan and Schmied, Christian-Marcel and Pr\"ufer, Maximilian and Kunkel, Philipp and Strohmaier, Robin and Strobel, Helmut and Gasenzer, Thomas and Kevrekidis, Panayotis G. and Oberthaler, Markus K.},
  journal = {Phys. Rev. Lett.},
  volume = {125},
  issue = {17},
  pages = {170401},
  numpages = {5},
  year = {2020},
  month = {Oct},
  publisher = {American Physical Society},
  doi = {10.1103/PhysRevLett.125.170401},
  url = {https://link.aps.org/doi/10.1103/PhysRevLett.125.170401}
}

@article{Heimburg:05,
ISSN = {00278424},
URL = {http://www.jstor.org/stable/3376057},
author = {Thomas Heimburg and Andrew D. Jackson and Gordon A. Baym},
journal = {Proceedings of the National Academy of Sciences of the United States of America},
number = {28},
pages = {9790--9795},
publisher = {National Academy of Sciences},
title = {On Soliton Propagation in Biomembranes and Nerves},
urldate = {2026-05-06},
volume = {102},
year = {2005}
}

@article{Veenstra:25,
  title = {Nonreciprocal Breathing Solitons},
  author = {Veenstra, Jonas and Gamayun, Oleksandr and Brandenbourger, Martin and van Gorp, Freek and Terwisscha-Dekker, Hans and Caux, Jean-S\'ebastien and Coulais, Corentin},
  journal = {Phys. Rev. X},
  volume = {15},
  issue = {3},
  pages = {031045},
  numpages = {20},
  year = {2025},
  month = {Aug},
  publisher = {American Physical Society},
  doi = {10.1103/nrv2-9h8z},
  url = {https://link.aps.org/doi/10.1103/nrv2-9h8z}
}

@article{Baker:25,
  title = {Observation of Jones-Roberts Solitons in a Paraxial Quantum Fluid of Light},
  author = {Baker-Rasooli, Myrann and Aladjidi, Tangui and Krause, Nils A. and Bradley, Ashton S. and Glorieux, Quentin},
  journal = {Phys. Rev. Lett.},
  volume = {134},
  issue = {23},
  pages = {233401},
  numpages = {8},
  year = {2025},
  month = {Jun},
  publisher = {American Physical Society},
  doi = {10.1103/PhysRevLett.134.233401},
  url = {https://link.aps.org/doi/10.1103/PhysRevLett.134.233401}
}

@article{Dieli:26,
  title = {Observation of Lump Solitons},
  author = {Dieli, Ludovica and Pierangeli, Davide and Baronio, Fabio and Trillo, Stefano and Conti, Claudio},
  journal = {Phys. Rev. Lett.},
  volume = {136},
  issue = {5},
  pages = {053804},
  numpages = {10},
  year = {2026},
  month = {Feb},
  publisher = {American Physical Society},
  doi = {10.1103/ggbs-y21w},
  url = {https://link.aps.org/doi/10.1103/ggbs-y21w}
}

@article{Rabec:25,
doi = {10.1038/s41567-025-02970-1},
url = {https://doi.org/10.1038/s41567-025-02970-1},
journal = {Nature Physics},
year = {2025},
month = {aug},
volume = {21},
number = {10},
pages = {1541},
author = {Rabec, F. and Chauveau, G. and Brochier, G. and Nascimbene, S. and Dalibard, J. and Beugnon, J.},
title = {Bloch oscillations of a soliton in a one-dimensional quantum fluid}
}

@article{Carbone:16,
doi = {10.1209/0295-5075/113/30003},
url = {https://doi.org/10.1209/0295-5075/113/30003},
year = {2016},
month = {feb},
publisher = {EDP Sciences, IOP Publishing and Società Italiana di Fisica},
volume = {113},
number = {3},
pages = {30003},
author = {Carbone, F. and Dutykh, D. and El, G. A.},
title = {Macroscopic dynamics of incoherent soliton ensembles: Soliton gas kinetics and direct numerical modelling},
journal = {Europhysics Letters}
}

@article{Novkoski:22,
doi = {10.1209/0295-5075/ac8a12},
url = {https://doi.org/10.1209/0295-5075/ac8a12},
year = {2022},
month = {sep},
publisher = {EDP Sciences, IOP Publishing and Società Italiana di Fisica},
volume = {139},
number = {5},
pages = {53003},
author = {Novkoski, Filip and Pham, Chi-Tuong and Falcon, Eric},
title = {Experimental observation of periodic Korteweg-de Vries solitons along a
torus of fluid},
journal = {Europhysics Letters},
}

@article{Suret:24,
  title = {Soliton gas: Theory, numerics, and experiments},
  author = {Suret, Pierre and Randoux, Stephane and Gelash, Andrey and Agafontsev, Dmitry and Doyon, Benjamin and El, Gennady},
  journal = {Phys. Rev. E},
  volume = {109},
  issue = {6},
  pages = {061001},
  numpages = {35},
  year = {2024},
  month = {Jun},
  publisher = {American Physical Society},
  doi = {10.1103/PhysRevE.109.061001},
  url = {https://link.aps.org/doi/10.1103/PhysRevE.109.061001}
}

@article{Fache:25,
  title = {Dissipation-Driven Emergence of a Soliton Condensate in a Nonlinear Electrical Transmission Line},
  author = {Fache, Loic and Damart, Herv\'e and Copie, Francois and Bonnemain, Thibault and Congy, Thibault and Roberti, Giacomo and Suret, Pierre and El, Gennady and Randoux, St\'ephane},
  journal = {Phys. Rev. Lett.},
  volume = {134},
  issue = {14},
  pages = {147201},
  numpages = {6},
  year = {2025},
  month = {Apr},
  publisher = {American Physical Society},
  doi = {10.1103/PhysRevLett.134.147201},
  url = {https://link.aps.org/doi/10.1103/PhysRevLett.134.147201}
}

@article{EL:16,
title = {Dispersive shock waves and modulation theory},                                      
journal = {Physica D: Nonlinear Phenomena},
volume = {333},
pages = {11-65},
year = {2016},
note = {Dispersive Hydrodynamics},
issn = {0167-2789},
doi = {https://doi.org/10.1016/j.physd.2016.04.006},
url = {https://www.sciencedirect.com/science/article/pii/S0167278916301580},
author = {G.A. El and M.A. Hoefer},                }

@ARTICLE{Zakharov:72,
   author = {V. E. Zakharov and A. B. Shabat},
   title = {Exact theory of two-dimensional self-focusing 
and one-dimensional self-modulation of waves in nonlinear media},
   journal = {Sov. Phys.--JETP},
   volume = {34},
   pages = {62-69},
   year = {1972}
}

@article{Zabusky:65,
  title = {Interaction of "Solitons" in a Collisionless Plasma and the Recurrence of Initial States},
  author = {Zabusky, N. J. and Kruskal, M. D.},
  journal = {Phys. Rev. Lett.},
  volume = {15},
  issue = {6},
  pages = {240--243},
  numpages = {0},
  year = {1965},
  month = {Aug},
  publisher = {American Physical Society},
  doi = {10.1103/PhysRevLett.15.240},
  OPTurl = {https://link.aps.org/doi/10.1103/PhysRevLett.15.240}
}

@article{Gardner:67,
  title = {Method for Solving the Korteweg-deVries Equation},
  author = {Gardner, Clifford S. and Greene, John M. and Kruskal, Martin D. and Miura, Robert M.},
  journal = {Phys. Rev. Lett.},
  volume = {19},
  issue = {19},
  pages = {1095--1097},
  numpages = {0},
  year = {1967},
  month = {Nov},
  publisher = {American Physical Society},
  doi = {10.1103/PhysRevLett.19.1095},
  url = {https://link.aps.org/doi/10.1103/PhysRevLett.19.1095}
}

@article{Ablowitz:73,
  title = {Nonlinear-Evolution Equations of Physical Significance},
  author = {Ablowitz, Mark J. and Kaup, David J. and Newell, Alan C. and Segur, Harvey},
  journal = {Phys. Rev. Lett.},
  volume = {31},
  issue = {2},
  pages = {125--127},
  numpages = {0},
  year = {1973},
  month = {Jul},
  publisher = {American Physical Society},
  doi = {10.1103/PhysRevLett.31.125},
  OPTurl = {https://link.aps.org/doi/10.1103/PhysRevLett.31.125}
}

@article{Copie:23,
title = {Space–time observation of the dynamics of soliton collisions in a recirculating optical fiber loop},
journal = {Optics Communications},
volume = {545},
pages = {129647},
year = {2023},
issn = {0030-4018},
doi = {https://doi.org/10.1016/j.optcom.2023.129647},
url = {https://www.sciencedirect.com/science/article/pii/S0030401823003954},
author = {François Copie and Pierre Suret and Stéphane Randoux}
}

@book{Newell_book_solitons,
  title={Solitons in mathematics and physics},
  author={Newell, Alan C},
  year={1985},
  publisher={SIAM}
}

@article{Miura:76,
author = {Miura, Robert M.},
title = {The Korteweg–deVries Equation: A Survey of Results},
journal = {SIAM Review},
volume = {18},
number = {3},
pages = {412-459},
year = {1976},
doi = {10.1137/1018076},
URL = {https://doi.org/10.1137/1018076}
}

@book{Drazin_book,
  title={Solitons: an introduction},
  author={Drazin, Philip G and Johnson, Robin Stanley},
  volume={2},
  year={1989},
  publisher={Cambridge university press}
}

@article{Maiden:18,
  title = {Solitonic Dispersive Hydrodynamics: Theory and Observation},
  author = {Maiden, Michelle D. and Anderson, Dalton V. and Franco, Nevil A. and El, Gennady A. and Hoefer, Mark A.},
  journal = {Phys. Rev. Lett.},
  volume = {120},
  issue = {14},
  pages = {144101},
  numpages = {5},
  year = {2018},
  month = {Apr},
  publisher = {American Physical Society},
  doi = {10.1103/PhysRevLett.120.144101},
  url = {https://link.aps.org/doi/10.1103/PhysRevLett.120.144101}
}

@article{GEl:05,
  title = {Kinetic Equation for a Dense Soliton Gas},
  author = {El, G. A. and Kamchatnov, A. M.},
  journal = {Phys. Rev. Lett.},
  volume = {95},
  issue = {20},
  pages = {204101},
  numpages = {4},
  year = {2005},
  month = {Nov},
  publisher = {American Physical Society},
  doi = {10.1103/PhysRevLett.95.204101},
  OPTurl = {https://link.aps.org/doi/10.1103/PhysRevLett.95.204101}
}

@article{Redor:19,
  title = {Experimental Evidence of a Hydrodynamic Soliton Gas},
  author = {Redor, Ivan and Barth\'elemy, Eric and Michallet, Herv\'e and Onorato, Miguel and Mordant, Nicolas},
  journal = {Phys. Rev. Lett.},
  volume = {122},
  issue = {21},
  pages = {214502},
  numpages = {6},
  year = {2019},
  month = {May},
  publisher = {American Physical Society},
  doi = {10.1103/PhysRevLett.122.214502},
  OPTurl = {https://link.aps.org/doi/10.1103/PhysRevLett.122.214502}
}

@article{Suret:20,
  title = {Nonlinear Spectral Synthesis of Soliton Gas in Deep-Water Surface Gravity Waves},
  author = {Suret, Pierre and Tikan, Alexey and Bonnefoy, F\'elicien and Copie, Fran\ifmmode \mbox{\c{c}}\else \c{c}\fi{}ois and Ducrozet, Guillaume and Gelash, Andrey and Prabhudesai, Gaurav and Michel, Guillaume and Cazaubiel, Annette and Falcon, Eric and El, Gennady and Randoux, St\'ephane},
  journal = {Phys. Rev. Lett.},
  volume = {125},
  issue = {26},
  pages = {264101},
  numpages = {6},
  year = {2020},
  month = {Dec},
  publisher = {American Physical Society},
  doi = {10.1103/PhysRevLett.125.264101},
  url = {https://link.aps.org/doi/10.1103/PhysRevLett.125.264101}
}

@article{Fache:24,
  title = {Interaction of soliton gases in deep-water surface gravity waves},
  author = {Fache, Loic and Bonnefoy, F\'elicien and Ducrozet, Guillaume and Copie, Francois and Novkoski, Filip and Ricard, Guillaume and Roberti, Giacomo and Falcon, Eric and Suret, Pierre and El, Gennady and Randoux, St\'ephane},
  journal = {Phys. Rev. E},
  volume = {109},
  issue = {3},
  pages = {034207},
  numpages = {13},
  year = {2024},
  month = {Mar},
  publisher = {American Physical Society},
  doi = {10.1103/PhysRevE.109.034207},
  url = {https://link.aps.org/doi/10.1103/PhysRevE.109.034207}
}

@article{GEl:21,
doi = {10.1088/1742-5468/ac0f6d},
url = {https://dx.doi.org/10.1088/1742-5468/ac0f6d},
year = {2021},
month = {nov},
publisher = {IOP Publishing and SISSA},
volume = {2021},
number = {11},
pages = {114001},
author = {Gennady A El},
title = {Soliton gas in integrable dispersive hydrodynamics},
journal = {Journal of Statistical Mechanics: Theory and Experiment}
}

@article{GEl:20,
  title={Spectral theory of soliton and breather gases for the focusing nonlinear Schr\"odinger equation},
  author={Gennady El and Alexander Tovbis},
  journal={Phys. Rev. E},
  volume={101},
  pages= {052207},
  year={2020},
}

@book{Novikov_book,
      author        = "Novikov, Sergei Petrovich and Manakov, S V and Pitaevskii, Lev Petrovich and Zakharov, Vladimir E",
      title         = "{Theory of solitons: the inverse scattering method}",
      publisher     = "Springer Science Business Media",
      OPTaddress       = "New York, NY",
      OPTseries        = "Contemporary Soviet mathematics",
      year          = "1984",
      OPTurl           = "https://cds.cern.ch/record/101022",
      OPTnote          = "Transl. of : Teoriia solitonov",
}

@book{yang2010nonlinear,
  title={Nonlinear Waves in Integrable and Non-integrable Systems},
  author={Yang, J.},
  OPTisbn={9780898717051},
  lccn={2010020210},
  series={Mathematical Modeling and Computation},
  OPTurl={https://books.google.fr/books?id=ACbbjcRQQvUC},
  year={2010},
  publisher={Society for Industrial and Applied Mathematics}
}

\section*{End Matter}

 \para{Recovering the KdV scattering shift from the interaction of two solitons} An important feature of soliton interactions is the scattering shift that remains after the collision. In the KdV equation, two interacting solitons undergo a non-trivial deformation during the collision but, at large times, the only persistent effect is this asymptotic shift, which plays a central role in SG theory and GHD. Although this effect is expected to be encoded in our construction, it is not immediately apparent from the definitions of the velocity field in \eqref{eq:VelField} and \eqref{eq:VelFieldDarDel}. We therefore consider the simplest two-soliton interaction and show that the scattering shift naturally emerges from the characteristics of the continuity equation \eqref{eq:ContEqEigen}.

Consider two solitons of parameters $\e_1$ and $\e_2$, such that $\e_1>\e_2$. When it comes to the interaction of two solitons,  the Darboux-deleted potential $\tilde u$ and the characteristic equations \eqref{eq:Charac} simplify to
\begin{equation}\label{eq:charac2sol}
\begin{aligned}
    &\dot x_1 = 4\e_1^2 + 4 \e_{2}^2\sech^2\left[\e_{2}(x_1-4\e_{2}^2t-\xi_{2})\right] \, , \\
    &\dot x_2 = 4\e_2^2 - 4 \e_{1}^2{\rm csch}^2\left[\e_{1}(x_2-4\e_{1}^2t-\xi_{1})\right] \, .
\end{aligned}
\end{equation}
At the height of interaction, when $x_i = 4\e_{3-i}t + \xi_{3-i}$, soliton 1 adds the bare velocity of soliton 2 to its own, while the velocity of 2 becomes singular. This singularity is what causes the temporary ``splitting'' of the slower soliton observable in Fig.~\ref{fig:SolInt}(e).

\begin{figure}[ht]
	\includegraphics[width=\linewidth]{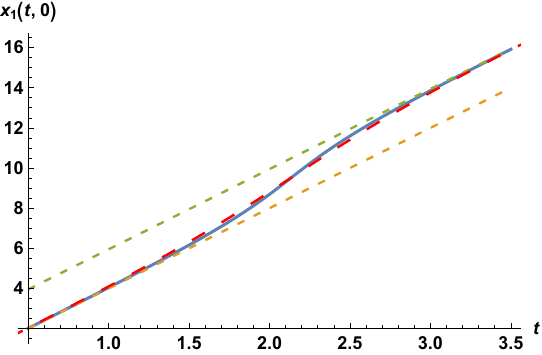} 
\caption{Characteristic trajectory (blue curve) starting at initial position $\bar x_1 = 0$ computed for solitons with parameters $\e_1 = 1$, $\e_2=0.75$, $\xi_2 = 4.7$. The orange dotted curve corresponds to the free trajectory $x = 4\e_1^2 t$ starting at the initial position $\bar x_1 =0$; the green dotted curve represents the same trajectory shifted by $\phi(\e_1,\e_2)$. The dashed red curve corresponds to the approximation \eqref{eq:characAp_2sol}.}
\label{fig:Cahrac_2sol}
\end{figure}

To improve readability, we now focus on the characteristic curve $x_1(t,\bar x_1)$, the characteristic $x_2(t,\bar x_2)$ can be treated in a similar fashion. Under the change of variable $y_1(t,\bar x_1) = \e_{2}\left[x_1(t,\bar x_1)-4\e_{2}^2t-\xi_{2}\right]$, the above ODE becomes separable
\begin{equation}\label{eq:ODEY}
\begin{aligned}
    \dot y_1 = \e_{2}\dot x_1 - 4 \e_{2}^3  = 4\e_2  \left[\e_1^2 - \e_2^2 \tanh^2(y_1)\right] \, .
\end{aligned} 
\end{equation}
This equation can easily be integrated and yields the following implicit expression of $y_1$
\begin{equation}\label{eq:ImplicitY}
    y_1(t) - F(y_1(t))  = 4\e_2 t(\e_1^2-\e_2^2)+ y_1(0) - F(y_1(0)) \, ,
\end{equation}
with the function $F(s) =   \frac{\e_2}{\e_1}\atanh\left[\frac{\e_2}{\e_1}\tanh(s)\right]$. Although it seems difficult to find an explicit solution of Eq.~\eqref{eq:ImplicitY}, it is possible to study its long time asymptotics. Recalling $\e_1>\e_2$, as $t\to\pm\infty$ we have that $(x_1(t,\bar x_1) -4\e_2^2t)\to \pm \infty$ such that
\begin{equation}\label{eq:lim}
\begin{aligned}
   \lim_{t\to\pm\infty}\atanh\left[\frac{\e_2}{\e_1}\tanh(y_1(t,\bar x_1))\right] = \pm\frac{1}{2}\log\left|\frac{\e_1+\e_2}{\e_1-\e_2}\right| \; .
\end{aligned}
\end{equation}
Given the definition of $y_1$, Eqs.~\eqref{eq:ImplicitY}-\eqref{eq:lim} yield
\begin{equation}
\begin{aligned}
   & \lim_{t\to\infty}\left[x_1(t,\bar x_1) - 4\e_1^2 t\right] - \lim_{t\to-\infty}\left[x_1(t,\bar x_1) - 4\e_1^2 t\right] \\
   &= \frac{1}{\e_1}\log\left|\frac{\e_1+\e_2}{\e_1-\e_2}\right| =\phi(\e_1,\e_2) \, , 
\end{aligned}
\end{equation}
and we indeed recover the scattering shift of KdV solitons \cite{novikov1984theory}, which is naturally embedded in the characteristics of the continuity equation \eqref{eq:ContEqEigen}. This is  confirmed through numerical integration of Eq.~\eqref{eq:charac2sol} as shown in Fig.~\ref{fig:Cahrac_2sol}. Making use of this asymptotic behavior, a somewhat reasonable approximation of $x_1(t,\bar x_1)$ can be devised (also illustrated in Fig.~\ref{fig:Cahrac_2sol})
\begin{equation}\label{eq:characAp_2sol}
\begin{aligned}
    &x_1(t,\bar x_1) \approx \bar x_1+4\e_1^2t + \frac{\phi(\e_1,\e_2)}{2}\left\{1 +\vphantom{\frac{\e_1+\e_2}{\e_1-\e_2}} \right.\\
    &\left.\tanh\left[\e_2\left(4t(\e_1^2-\e_2^2) - \xi_2+\bar x_1 +\frac{\phi(\e_1,\e_2)}{2}\right)\right]\right\}\, .
\end{aligned}
\end{equation}

Along the curve $x_1(t,\bar x_1)$ the squared eigenfunction evolves according to the separable ODE
    \begin{equation}
    \begin{aligned}
        &\dot Q_1(x_1(t,\bar x_1),t) = -4Q_1(x_1(t,\bar x_1),t)\\
        &\times \partial_{x_1} \left[\e_1^2+\e_2^2\sech^2\left[\e_2(x_1(t,\bar x_1)-4\e_2^2t-\xi_2)\right]\right] \, , 
    \end{aligned}
\end{equation}
which can be integrated as
  \begin{equation}\label{eq:EigenEvol_2sol}
  \begin{aligned}
    \log&\frac{Q_1(x_1(t,\bar x_1),t)}{\bar Q_1(\bar x_1)}=- 4\e_2^3\int_0^t \dd \tau\, \partial_{y_1}\sech^2\left[y_1(\tau,\bar x_1)\right] \, . 
    \end{aligned}
\end{equation}  
Note how $Q_1$ decreases during the interaction: the inhomogeneous velocity field deforms the soliton which widens as part of it is accelerated; conservation of its mass implies that its amplitude must decrease (see Fig.~\ref{fig:SolInt}). Although we do not have an explicit solution for the characteristic curve $x_1(t,\bar x_1)$ we may still evaluate the integral \eqref{eq:EigenEvol_2sol}. Using the differential equation \eqref{eq:ODEY}, we may perform the following change of variable
\begin{equation}
    \log\frac{Q_1(x_1(t,\bar x_1),t)}{\bar Q_1(\bar x_1)} = 2\e_2^2\int_{y_1(0,\bar x_1)}^{y_1(t,\bar x_1)} \dd y \,\frac{\sech^2\left(y\right)\tanh\left(y\right)}{\e_1^2-\e_2^2 \tanh^2(y)} \, .
\end{equation}
We may then perform a second change of variable $u = \tanh(y)$ so that
\begin{equation}\label{eq:ExactEigenEvol_2sol}
\begin{aligned}
    &\log\frac{Q_1(x_1(t,\bar x_1),t)}{\bar Q_1(\bar x_1)} = 2\e_2^2\int_{u_1(0,\bar x_1)}^{u_1(t,\bar x_1)} \dd u \,\frac{u}{\e_1^2-\e_2^2 u^2} \\
    &= \log\left(\frac{\e_1^2-\e_2^2\tanh^2\left[\e_2(\bar x_1-4\e_2^2t-\xi_2)\right]}{\e_1^2-\e_2^2\tanh^2\left[\e_2(x_1(t,\bar x_1)-4\e_2^2t-\xi_2)\right]}\right) \, .
\end{aligned}
\end{equation}
For $|x_1(t,\bar x_1) - 4\e_2^2 t-\xi_2|\gg1$ we have
\begin{equation}
    \log\frac{Q_1(x_1(t,\bar x_1),t)}{\bar Q_1(\bar x_1)} \approx 0 \ \Rightarrow \ Q_1(x_1(t,\bar x_1),t) \approx \bar Q_1(\bar x_1) \, .
\end{equation}
While $Q_1$ deforms during the interaction, it then recovers its initial shape and is simply transported along the characteristic curves, which is consistent with well-known soliton phenomenology. With that, the evolution of the squared eigenfunction is fully characterized in this simple case.

\para{Rarefied soliton gas} What we discussed in the previous section can be readily extended to account for the rarefied SG regime described in \cite{zakharov1971kinetic}. In the well-separated regime, if $\e_m>\e_n$ for $m<n$
\begin{equation}\label{eq:rarefied}
\begin{aligned}
    v_m(x,t)
\approx
4\e_m^2
&+
4\sum_{n > m}
\e_n^2
\sech^2\!\left[\e_n(x-4\e_n^2 t-\xi_n)\right] \\
&-4\sum_{n < m}
\e_n^2
{\rm csch}^2\!\left[\e_n(x-4\e_n^2 t-\xi_n)\right]\; .
\end{aligned}
\end{equation}
Assuming the solitons are sufficiently well separated so that the $m-$th will only interact with one soliton at a time, it is possible to partition the $(x,t)-$plane  into $(N-1)$  regions $\{R_n\}_{n=1, \ n\neq m}^N$, with $\cup_{n=1, \ n\neq m}^N R_n = \mathbb R^2$, in which (e.g. if $n>m$) the velocity field can be further approximated by
\begin{equation}
    v_m(x,t) \approx 4\e_m^2 + 4 \e_n^2\sech^2\left[\e_n(x-4\e_n^2 t-\xi_n)\right] \; .
\end{equation}
 If so, using the previously discussed methods, resolution of the continuity equation \eqref{eq:ContEqEigen} is immediate. In particular, characteristics can be approximated by 
\begin{equation}
\hspace{-15pt}\begin{aligned}
   &\quad x_m(t,\bar x_m) \approx \bar x_m+4\e_m^2t+\sum_{n > m} \frac{\phi(\e_m,\e_n)}{2}\left\{1 +\vphantom{\frac{1}{2}}\right. \\
    &\quad \left. \tanh\left[\e_n\left(4t(\e_m^2-\e_n^2) - \xi_n +\bar x_m+\frac{\phi(\e_m,\e_n)}{2}\right)\right]\right\}\, \\
    &-\sum_{n < m}  \phi(\e_m,\e_n) \Theta\left[4t(\e_m^2-\e_n^2) - \xi_n +\bar x_m+\frac{\phi(\e_m,\e_n)}{2}\right].
\end{aligned}
\end{equation}
Those can further be approximated by trajectories of the \textit{semiclassical Bethe systems} introduced in \cite{doyon2024new} and further studied in \cite{doyon2026generalised}.

\begin{figure}[ht]
	\includegraphics[width=\linewidth]{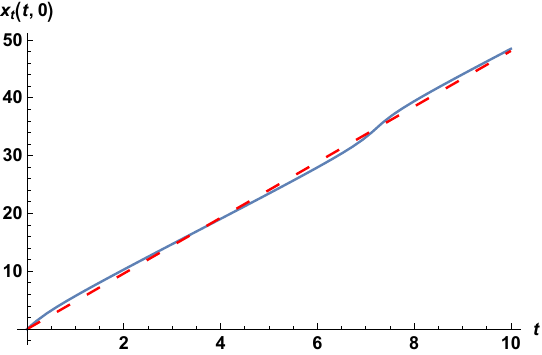}  
\caption{Characteristic trajectory of a soliton ($\eta_t = 1$, $\bar x_t=0$) within a genus 1 condensate ($a = .1$, $b=.92$, $c=1$), obtained by solving \eqref{eq:yGenus1} (full blue line) compared to the trajectory $x = v^{{\rm eff}}(\eta) t$ (dashed red line) where $v^{{\rm eff}}$ is the effective velocity of a tracer obtained in \cite{congy2023dispersive} by solving Eq.~\eqref{eq:Veff}.}
\label{fig:CharacGenus1}
\end{figure}

\para{Tracer in genus-1 soliton condensate} The so-called \textit{genus-1 soliton condensate} (see \cite{congy2023dispersive} for a detailed description, including the associated DOS) corresponds to a deterministic wave-field
\begin{equation}
    u_{c}(x,t) = a + b + c + 2(c-a) {\rm dn}^2\left[\theta(x,t);m \right]\, ,
\end{equation}
where $0<a<b<c$ are positive constants, $m=(b-a)/(c-a)$, ${\rm dn}$ is the Jacobi elliptic function, and $\theta(x,t) = [x-2(a+b+c)t]\sqrt{c-a}$. Via the definition of the velocity field \eqref{eq:VelFieldDarDel}, the characteristics $x_t(t,\bar x_t)$ of a tracer soliton of parameter $\e_t\geq c$ are given by a separable ODE in terms of $y_t(t,\bar x_t) = \theta(x_t(t,\bar x_t), t)$
\begin{equation}\label{eq:yGenus1}
    \dot y_t = 4\sqrt{c-a}\left[\e_t^2-c+(c-a){\rm dn}^2(y_t|m)\right] \, .
\end{equation}
Using the identity ${\rm dn}^2(x|m)=1-m\,{\rm sn}^2(x|m)$ this ODE can be integrated into the implicit relation (c.f. Eq.~(400.01) of \cite{byrd2013handbook})
\begin{equation}
    \frac{1}{\e^2_t-a}\Pi\left[\frac{(c-a)m}{\e_t^2-a};{\rm A}\left(y_1(t,\bar x_1)|m\right)\biggl|m\right] = 4t\sqrt{c-a} + C \, ,
\end{equation}
where $\Pi$ is the incomplete elliptic integral of the third kind, A the Jacobi amplitude and $C$ an integration constant. In Fig.~\ref{fig:CharacGenus1} we compare the exact characteristic trajectory obtained by solving Eq.~\eqref{eq:yGenus1} to a ballistic trajectory moving with the effective velocity $v^{{\rm eff}}(\eta)$ obtained in \cite{congy2023dispersive} by solving Eq.~\eqref{eq:Veff}. Here, it is clear that $v^{{\rm eff}}(\eta)$ corresponds to a temporal average of the characteristic velocity which lends credence to our hypothesis \eqref{eq:VeffMicro}. Furthermore, this shows our method provides a way of recovering results analogous to that of \cite{girotti2023soliton} while being simpler, more straightforward and more general since we do not need the amplitude of the tracer to be large compared to that of the condensate.

\noindent Remark: the \textit{genus-0 condensate} corresponding to a constant field \cite{congy2023dispersive}, $u_c = \alpha$, we trivially recover the results of \cite{congy2026exactly} regarding the velocity of a tracer, $\dot x_t = 4\e_t^2+2\alpha$.

\end{document}